\documentclass[12pt]{article}
\usepackage{bezier}
\usepackage{amssymb}
\usepackage{epsfig}

\newcommand{\nc}{\newcommand}
\nc{\be}{\begin{equation}}
\nc{\ee}{\end{equation}}
\nc{\bea}{\begin{eqnarray}}
\nc{\eea}{\end{eqnarray}}

\nc{\eqn}[1]{{(\ref{#1})}}
\nc{\cA}{{\cal A}}
\nc{\cB}{{\cal B}}
\nc{\cC}{{\cal C}}
\nc{\cD}{{\cal D}}
\nc{\cE}{{\cal E}}
\nc{\cF}{{\cal F}}
\nc{\cG}{{\cal G}}
\nc{\cH}{{\cal H}}
\nc{\cI}{{\cal I}}
\nc{\cJ}{{\cal J}}
\nc{\cK}{{\cal K}}
\nc{\cL}{{\cal L}}
\nc{\cM}{{\cal M}}
\nc{\cN}{{\cal N}}
\nc{\cO}{{\cal O}}
\nc{\cP}{{\cal P}}
\nc{\cQ}{{\cal Q}}
\nc{\cR}{{\cal R}}
\nc{\cS}{{\cal S}}
\nc{\cT}{{\cal T}}
\nc{\cU}{{\cal U}}
\nc{\cV}{{\cal V}}
\nc{\cW}{{\cal W}}
\nc{\cX}{{\cal X}}
\nc{\cY}{{\cal Y}}
\nc{\cZ}{{\cal Z}}
\nc{\simo}[1]{{\stackrel{#1}{\simeq}}}
\nc{\geqo}[1]{{\stackrel{#1}{\geq}}}
\nc{\geo}[1]{{\stackrel{#1}{>}}}
\nc{\guo}[1]{{\stackrel{#1}{\succ}}}
\nc{\rbo}{\raisebox}
\nc{\RR} {\rangle \! \rangle}
\nc{\LL} {\langle \! \langle}
\nc{\rmi}[1]{{\mbox{\small #1}}}
\nc{\eq}{eq.~}
\nc{\nr}[1]{(\ref{#1})}
\nc{\ul}{\underline}
\nc{\mc}{\multicolumn}
\nc{\todo}[1]{\par\noindent{\bf $\rightarrow$ #1}}

\nc{\cu}{{\cal u}}

\title{
  \begin{flushright} {\small $\begin{array}{ l } 
\mbox{HD--THEP--00--16} \\
    \end{array} $}
 \end{flushright}
\vskip 2cm
Universality of the Axial Anomaly in Lattice QCD}

\author{M. Frewer and H.~J.~Rothe
                 \\ \\Institut
        f\"ur Theoretische Physik,\\
        Universit\"at Heidelberg, \\
        Philosophenweg 16, \\
        D-69120 Heidelberg, Germany}

\begin{document}
\maketitle

\begin{abstract}

We prove that lattice QCD generates the axial anomaly in 
the continuum limit under very general conditions on the lattice action, 
which includes the case of Ginsparg-Wilson fermions.  
The ingredients going into the proof are gauge invariance, locality of the 
Dirac operator, absence of fermion doubling, the general form of the 
lattice Ward identity, and the power counting theorem of Reisz. The results generalize in an obvious way to $SU(N)$ lattice gauge theories.
 
\end{abstract}
\vfill
\eject
\noindent
{\bf 1. Introduction}
\vskip7pt
In the continuum formulation of QCD (or QED) it is well known that different 
gauge invariant regularization schemes all yield the same expression for the axial anomaly. A non-perturbative regularization is provided by the lattice. Any candidate for a lattice discretization of QED or QCD should reproduce the correct axial anomaly in the continuum limit. 
For lattice QED with Wilson fermions the axial anomaly has been first studied by Karsten and Smit \cite{Karsten}. These authors showed that the origin of the anomaly was an irrelevant term in the lattice Ward identity. In \cite{Sadooghi} it was shown, using the small-a-expansion scheme of ref. \cite{Wetzel}, that in the limit of vanishing lattice spacing $a$ this naively irrelevant contribution is necessarily given by the $D\to 4$ limit of the dimensionally regulated continuum triangle graph. That the correct anomaly is in fact generated for any  lattice discretization of the QED action satisfying a very general set of conditions, which include the case of Ginsparg-Wilson fermions, has been recently shown in 
\cite{Reisz-Rothe}. It is the purpose of this paper to generalize this result to the 
non-abelian case.  Although for concreteness sake we shall consider the case of QCD, it will be evident 
from our discussion that our proof goes through for any $SU(N)$ gauge theory.

The paper is organized as follows. In the following section we discuss the general form of the lattice axial vector Ward identity, which is one of the  
important ingredients which will be needed for the proof. As we shall see, its precise structure, which depends on the particular discretization of the lattice action, need not be known, but only very general properties thereof. In section 
3 we then state the theorem we want to prove, and discuss the consequences of 
the various (very general) conditions on the action required by the theorem. 
The proof of the theorem is presented in section 4.
\vskip7pt
%\vfill
%\eject
\noindent
{\bf 2. Ward Identity} 
\vskip7pt
Consider the following general form of the lattice action for QCD
\be
S = S_G[U] + \sum_{x,y}{\bar\psi}(x)(D_U(x,y)+m)\psi(x)\ , 
\ee
where $S_G[U]$ is the standard Wilson gauge field action and $D_U(x,y)$ is the 
Dirac operator (a matrix in Dirac-spin and colour space). $U$ denotes collectively 
the link variables $U_\mu(x) = \exp[igaA_\mu(x)]$, with $a$ the lattice spacing,  and $\psi$, $\bar\psi$ are three component colour Dirac fields. $\sum_x = \sum_n a^4$, where $n$ labels the lattice sites. The action is assumed to be gauge invariant, and to possess the discrete symmetries of the continuum theory.
The Dirac operator has the following formal expansion in the gauge potentials
\be
D_U(x,y) = \sum_{n,\mu_i,a_i,x_i}\frac{1}{n!}D^{{(n)}a_1\cdot\cdot\cdot a_n}_{\mu_1\cdot
\cdot\cdot\mu_n}(x,y|x_1\cdot\cdot\cdot x_n)A^{a_1}_{\mu_1}(x_1)\cdot
\cdot\cdot A^{a_n}_{\mu_n}(x_n)\ .
\ee
It can always be decomposed into a chirally symmetric, and a chiral 
symmetry breaking (sb) part as follows,
\be
D_U(x,y) = D_U(x,y)_{sym}+ D_U(x,y)_{sb}\ ,
\ee
where
\be
D_U(x,y)_{sym} = \frac{1}{2}[D_U,\gamma_5]\gamma_5\  \ , 
\ee
\be   
D_U(x,y)_{sb} =\frac{1}{2}\{D_U,\gamma_5\}\gamma_5\ . 
\ee
Any candidate for a lattice action should possess the correct continuum limit. It therefore follows that for $a\to 0$, $D_U(x,y)_{sym} \to 
\gamma_\mu D_\mu[A]$, where $ D_\mu[A]$ is the covariant derivative, while 
$D_U(x,y)_{sb}$ vanishes in the continuum limit. 

Let $\delta S$ be the variation of the action induced by the (standard) 
colour singlet axial 
transformation
\bea
\delta\psi(x) &=& i\epsilon(x)\gamma_5\psi(x)\ .\cr 
\delta{\bar\psi}(x) &=& i\epsilon(x){\bar\psi}(x)\gamma_5\ . 
\eea
Because the measure is invariant under this transformation, it follows 
that
\be
<\delta S>_U = 0 \ ,
\ee
where $<\delta S>_U$ denotes the expectation value of 
$\delta S$ in the external background field $\{U_\mu(n)\}$. As a consequence of the Poincare Lemma on the lattice \cite{Poincare} the variation of the action 
can always be written in the form
\be
\delta S = i\sum_x \omega(x)\Big\lbrack-\partial^L_\mu j^5_\mu (x) + 
2m j_5(x) + \Delta (x)\Big\rbrack \ ,
\ee
where $j_5(x) = {\bar\psi}(x)\gamma_5\psi(x)$, $j^5_\mu(x)\to  {\bar\psi}(x)\gamma_\mu\gamma_5\psi(x)$ for $a\to 0$, and $\Delta(x)$ is 
an irrelevant operator vanishing in the continuum limit. ${\partial^L_\mu}$ is the (dimensioned) left lattice derivative.
Hence the Ward identity for the colour singlet axial vector current will be of the general form
\be
<{\partial}^L_\mu j^5_\mu(x)>_U = 2m<j_5(x)>_U + <\Delta(x)>_U \ .
\ee
The lattice definitions of the various (gauge invariant) operators in (9) are only unique up to terms vanishing in the naive continuum 
limit. Hence the choice of the left derivative in (9) is only a matter of 
convenience. It appears naturally, e.g., when Wilson fermions are considered. 

Let ${\cal O}(x)$ stand for any of the operators appearing in (9). Then  
$<{\cal O}(x)>_U$ has the following formal expansion in the gauge potentials
\be
<{\cal O}(x)>_U = \sum_{n\ge 2}\frac{1}{n!}\sum_{\{x_i\},\{\mu_i\},\{a_i\}}\Gamma^{({\cal O})a_1\cdot\cdot\cdot a_n}_{\mu_1\cdot\cdot\cdot\mu_n}(x|x_1,x_2,\cdot\cdot\cdot ,x_n)
A^{a_1}_{\mu_1}(x_1)\cdot\cdot\cdot A^{a_n}_{\mu_n}(x_n)\ ,
\ee
 where, because of the assumed symmetries of (1), the sum over $n$ starts with $n=2$. The correlation functions $\Gamma^{({\cal O})a_1\cdot\cdot\cdot a_n}_{\mu_1\cdot\cdot\cdot\mu_n}(x|x_1,x_2,\cdot\cdot\cdot, x_n)$ are symmetric 
under the exchange of any pair of collective labels $(x_i,\mu_i,a_i)$. Defining the Fourier transform of $\Gamma^{({\cal O})a_1\cdot\cdot\cdot a_n}_{\mu_1\cdot\cdot\cdot\mu_n}(x|x_1,x_2,\cdot\cdot\cdot ,x_n)
$ by
\be
\Gamma^{({\cal O})a_1\cdot\cdot\cdot a_n}_{\mu_1\cdot\cdot\cdot\mu_n}(x|x_1,\cdot\cdot\cdot ,x_n) = \int^{\frac{\pi}{a}}_{-\frac{\pi}{a}}\frac{d^4q}{(2\pi)^4}e^{-iq\cdot x}
\prod^{n}_{i=1}\frac{d^4k_i}{(2\pi)^4}e^{ik_i\cdot x_i}{\hat\Gamma}^{({\cal O})a_1\cdot\cdot\cdot a_n}_{\mu_1\cdot\cdot\cdot\mu_n}(q|k_1,\cdot\cdot\cdot,k_n)\ ,  
\ee
where, by translational invariance,
\be
{\hat\Gamma}^{({\cal O})a_1\cdot\cdot\cdot a_n}_{\mu_1\cdot\cdot\cdot\mu_n}(q|k_1,\cdot\cdot\cdot,k_n) = 
\delta(q-\sum^n_{i=1} k_i){\tilde\Gamma}^{({\cal O})a_1\cdot\cdot\cdot a_n}_{\mu_1\cdot\cdot\cdot\mu_n}(k_1,\cdot\cdot\cdot,k_n)\ , 
\ee
the Ward identity (9) translates as follows to momentum space,
\bea
-i{\tilde q}_{\mu}{\tilde\Gamma}^{a_1\cdot\cdot\cdot a_n}_{5\mu;\mu_1\cdot\cdot\cdot\mu_n}(k_1,\cdot\cdot\cdot,k_n) &=& 
2m{\tilde\Gamma}^{a_1\cdot\cdot\cdot a_n}_{5;\mu_1\cdot\cdot\cdot\mu_n}
(k_1,\cdot\cdot\cdot,k_n)\cr\cr
 &+& {\tilde\Gamma}^{(\Delta)a_1\cdot\cdot\cdot a_n}_{\mu_1\cdot\cdot\cdot\mu_n}
(k_1,\cdot\cdot\cdot,k_n)\ ,
\eea
where
\be
{\tilde q}_\mu = e^{i\frac{q_\mu a}{2}}\frac{2}{a}\sin\frac{q_\mu a}{2}\ .
\ee
As we shall see further below, gauge invariance implies that every term in 
eq. (13) possesses a finite continuum limit. Any anomalous contribution 
to the divergence of the axial vector current must therefore have its origin in the 
naively irrelevant $\Delta$ term. If its continuum limit is universal, then 
so is the anomaly.
 
We now state the central theorem of this paper and then give the proof:
\vskip7pt
\noindent
{\bf 3. Theorem}
\vskip7pt
Any lattice discretization of the action (1) with the following properties: a) S has the correct continuum limit; 
b) S is gauge invariant; c) The Dirac operator is local and d) The free Dirac operator is invertible for nonvanishing momenta (no doublers), reproduces the axial anomaly in the continuum limit. 

Let us consider first in turn the consequences of b) to d). 
\vskip7pt
%\vfill
%\eject
\noindent
i) {\it Gauge Invariance}
\vskip7pt
\noindent
Gauge invariance tells us that if ${\cal O}(A,\psi,{\bar\psi})$ is a gauge invariant operator, then its external field expectation value satisfies  
\be
<{\cal O}(A^\omega,\psi,{\bar\psi})>_{A^\omega} =
<{\cal O}(A,\psi,{\bar\psi})>_A\ ,
\ee
where $A^\omega$ is the gauge transformed potential. On the lattice  
the variation of the gauge potentials induced by an infinitessimal gauge transformation is given by \cite{Kawai}
\be
\delta A^a_\mu(x) = [gf_{abc}A^b_\mu(x) -M^{-1}_{ac}(gaA_\mu(x))\partial^R_\mu]\epsilon^c(x)\ ,
\ee
where $f_{abc}$ are the structure constants of $SU(3)$, $\partial^R_\mu$ is 
the dimensioned right lattice derivative, and the matrix $M$ is given by 
\be
M(\phi(x)) = \frac{1-e^{-i\phi(x)}}{i\phi(x)}\ ,
\ee
with $\phi$ a colour matrix in the adjoint representation. A summation over repeated indices is 
understood. Up to ${\cal O}(\phi^4)$ we have that
\be
M^{-1}(\phi) = 1 + \frac{i}{2}\phi - \frac{1}{12}(\phi^2) + {\cal O}(\phi^4)\ .
\ee
Performing the transformation (16) in (10), one is led to an infinite set of gauge identities. Only the following two will however be 
required for the proof of the theorem:
\be 
\partial^L_{\mu_1}\Gamma^{({\cal O})a_1a_2}_{\mu_1\mu_2}(x|x_1x_2) = 0\ ,
\ee
\bea
\partial^L_{\mu_1}\Gamma^{({\cal O})a_1a_2a_3}_{\mu_1\mu_2\mu_3}(x|x_1x_2x_3) &=&
g\delta_{x_1x_2}f_{a_1a_2c}\Gamma^{({\cal O})ca_3}_{\mu_2\mu_3}(x|x_2x_3)\cr
&-& \frac{1}{2!}agf_{a_1a_2c}\partial^L_{\mu_2}\delta_{x_1x_2}
\Gamma^{({\cal O})ca_3}_{\mu_2\mu_3}(x|x_2x_3)\cr\cr
&+&(2\leftrightarrow 3)\ ,
\eea
where the left-derivatives acts on $x_1$. 
Written in momentum space (19) takes the form
\be
({\tilde k}^*_1)_{\mu_1}{\tilde\Gamma}^{({\cal O})a_1a_2}_{\mu_1\mu_2}(k_1,k_2) 
= 0\ ,
\ee
where $*$ denotes complex conjugation, and 
where ${\tilde k}_\mu$ is defined in an analogous way to (14). 
Because of Bose symmetry a corresponding statement holds for $({\tilde k}_1)_{\mu_1}$ 
replaced by  $({\tilde k}_2)_{\mu_2}$. The second identity (20) reads as 
follows in momentum space, 
\bea
i({\tilde k}^*_1)_{\mu_1}{\tilde\Gamma}^{({\cal O})a_1a_2a_3}_{\mu_1\mu_2\mu_3}(k_1,k_2,k_3)
&=&gf_{a_1a_2c}{\tilde\Gamma}^{({\cal O})ca_3}_{\mu_2\mu_3}(k_1+k_2,k_3)\cr\cr
&-& ia\frac{g}{2!}f_{a_1a_2c}({\tilde k_1}^*)_{\mu_2}{\tilde\Gamma}^{{\cal O}ca_3}_{\mu_2\mu_3}(k_1+k_2,k_3)\cr\cr
&+& (2\leftrightarrow 3)\ .
\eea

\vskip7pt
\noindent 
ii) {\it Locality of the Dirac operator}
\vskip7pt
The coefficient functions $D^{{(n)}a_1\cdot\cdot\cdot a_n}_{\mu_1\cdot
\cdot\cdot\mu_n}(x,y|x_1\cdot\cdot\cdot x_n)$ in (2) are assumed to vanish 
at least exponentially fast for large separations between any pair of lattice sites with a decay constant of the order of the inverse lattice spacing. This is trivially the case for Wilson fermions, where the Dirac operator connects only neighbouring lattice sites. It also holds for Ginsparg-Wilson fermions, for sufficiently smooth background gauge field configurations \cite{Jansen}. As a 
consequence the correlation functions ${\tilde\Gamma}^{({\cal  O})a_1\cdot\cdot\cdot a_n}_{\mu_1\cdot\cdot\cdot\mu_n}$ are analytic functions of the 
momenta around vanishing momenta. This fact, combined with eq. (21)  
(following from gauge invariance), implies that
\be
T_1{\tilde\Gamma}^{({\cal O})a_1a_2}_{\mu_1\mu_2}(k_1,k_2) = 0\ ,
\ee
where $T_1$ denotes the Taylor expansion around zero momenta  
up to first order.  

Consider next eq.(22) evaluated for $k_2=k_3 = 0$, and small $k_1$.
Because of (23) the RHS is of ${\cal O}(k^2_1)$. It therefore follows that
\be
{\tilde\Gamma}^{({\cal O})a_1a_2a_3}_{\mu_1\mu_2\mu_3}(0,0,0) = 0\ .
\ee
Relations (23) and (24) are weaker than those obtained from gauge invariance in the case of QED, where an analogous statement to (23) holds for vertex functions involving an arbitrary number of external photon lines. They nevertheless suffice to prove the theorem.
\vskip7pt
\noindent
iii) {\it The free propagator} $D^{(0)}(p)^{-1}$ {\it is free of doublers}
\vskip7pt
As a consequence the lattice power counting rules and the Reisz theorem \cite{Reisz1} applies. This theorem allows us to take the naive continuum limit of a 
Feynman integral, if the lattice degrees of divergence of all Zimmermann 
spaces are negative.
\vskip7pt
\noindent
{\bf 4. Proof of the theorem}
\vskip7pt
\noindent
We are now ready to prove the theorem. Our attention will be focused on the 
"irrelevant" contribution of the $\Delta$-term in (13), which should generate the anomaly in the continuum limit. 

Consider the axial vector Ward identity (13). By power counting the ultraviolet lattice degree of divergence (LDD) of ${\tilde\Gamma}^5_\mu$ and ${\tilde\Gamma}_5$ 
is given by $3-n$, where $n$ is the number of external gluon fields, while 
the LDD of ${\tilde\Gamma}^{(\Delta)}$ is $4-n$. Hence the LDD 
of Feynman integrals contributing to $\Gamma^{(\Delta)}$ is negative for graphs involving more than 4 external gauge fields. Their contributions thus vanish by the Reisz theorem \cite{Reisz1} in the continuum limit, since $\Delta$ is an irrelevant operator. We therefore only need to consider the correlation functions  
${\tilde\Gamma}^{(\Delta)a_1a_2}_{\mu_1\mu_2}(k_1,k_2)$, ${\tilde\Gamma}^{(\Delta)a_1a_2a_3}_{\mu_1\mu_2\mu_3}(k_1,k_2,k_3)$, and
${\tilde\Gamma}^{(\Delta)a_1a_2a_3a_4}_{\mu_1\mu_2\mu_3\mu_4}(k_1,k_2,k_3,k_4 )$ with $LDD = 2, 1$ and $0$, respectively. Let us decompose these vertex functions as follows
\bea
{\tilde\Gamma}^{(\Delta)a_1a_2}_{\mu_1\mu_2}(k_1,k_2) &=& 
(1-T_2){\tilde\Gamma}^{(\Delta)a_1a_2}_{\mu_1\mu_2}(k_1,k_2)\cr\cr
 &+& T_1{\tilde\Gamma}^{(\Delta)a_1a_2}_{\mu_1\mu_2}(k_1,k_2) + (T_2-T_1)
{\tilde\Gamma}^{(\Delta)a_1a_2}_{\mu_1\mu_2}(k_1,k_2)\ ,
\eea
\bea
{\tilde\Gamma}^{(\Delta)a_1a_2a_3}_{\mu_1\mu_2\mu_3}(k_1,k_2,k_3) &=&
(1-T_1){\tilde\Gamma}^{(\Delta)a_1a_2a_3}_{\mu_1\mu_2\mu_3}(k_1,k_2,k_3)
+ T_0{\tilde\Gamma}^{(\Delta)a_1a_2a_3}_{\mu_1\mu_2\mu_3}(k_1,k_2,k_3)\cr
&+& (T_1-T_0){\tilde\Gamma}^{(\Delta)a_1a_2a_3}_{\mu_1\mu_2\mu_3}(k_1,k_2,k_3)\ ,
\eea
and 
\bea
{\tilde\Gamma}^{(\Delta)a_1a_2a_3a_4}_{\mu_1\mu_2\mu_3\mu_4}(k_1,k_2,k_3,k_4 ) 
&=&
(1-T_0){\tilde\Gamma}^{(\Delta)a_1a_2a_3a_4}_{\mu_1\mu_2\mu_3\mu_4}(k_1,k_2,k_3,k_4 )\cr
&+&T_0{\tilde\Gamma}^{(\Delta)a_1a_2a_3a_4}_{\mu_1\mu_2\mu_3\mu_4}
(k_1,k_2,k_3,k_4 )\ ,
\eea 
where $T_n$ denotes the Taylor expansion around vanishing momenta up to 
n'th order. 
By the Reisz theorem the first term appearing on the RHS of (25) to (27), 
respectively, vanishes in the continuum limit, since it has negative LDD and 
$\Delta$ is an irrelevant operator. The respective second terms in (25) and (26) vanish by 
gauge invariance (cf. eqs (23) and (24)). Hence we conclude that
\be
\lim_{a\to 0}{\tilde\Gamma}^{(\Delta)a_1a_2}_{\mu_1\mu_2}(k_1,k_2) = 
\lim_{a\to 0} (T_2-T_1)
{\tilde\Gamma}^{(\Delta)a_1a_2}_{\mu_1\mu_2}(k_1,k_2)\ ,
\ee
\be
\lim_{a\to 0}{\tilde\Gamma}^{(\Delta)a_1a_2a_3}_{\mu_1\mu_2\mu_3}(k_1,k_2,k_3) =  \lim_{a\to 0}(T_1-T_0){\tilde\Gamma}^{(\Delta)a_1a_2a_3}_{\mu_1\mu_2\mu_3}(k_1,k_2,k_3)\ , 
\ee
\be
\lim_{a\to 0}{\tilde\Gamma}^{(\Delta)a_1a_2a_3a_4}_{\mu_1\mu_2\mu_3\mu_4}(k_1,k_2,k_3,k_4 ) = \lim_{a\to 0}T_0{\tilde\Gamma}^{(\Delta)a_1a_2a_3a_4}_{\mu_1\mu_2\mu_3\mu_4}(k_1,k_2,k_3,k_4 )\ .
\ee
As we now show the limits on the RHS are universal, i.e., they do not depend on the specific form of the $\Delta$ operator in (9).

Consider first the RHS of (25). Applying the operation 
$(T_2-T_1)$ to the axial Ward identity (13) with $n=2$, and making use of (23), following 
from gauge invariance, we conclude that
\be
(T_2-T_1){\tilde\Gamma}^{(\Delta)a_1a_2}_{\mu_1\mu_2}
(k_1,k_2) = -2m(T_2-T_1){\tilde\Gamma}^{a_1a_2}_{5;\mu_1\mu_2}
(k_1,k_2)\ .
\ee
The expression appearing on the RHS has negative LDD; hence its continuum limit 
is given, according to the Reisz theorem, by applying the $(T_2-T_1)$ operation to the integrand of the 
corresponding continuum Feynman integral, i.e. the triangle diagram.  
One then readily finds that (28) is given by
\be
\lim_{a\to 0}{\tilde\Gamma}^{(\Delta)a_1a_2}_{\mu_1\mu_2}
(k_1,k_2)  = -\frac{g^2}{4\pi^2}\delta_{a_1a_2}\epsilon_{\mu_1\mu_2\sigma\rho}(k_1)_\sigma 
(k_2)_\rho \ .
 \ee
Consider next the RHS of (29). Applying the 
$(T_1-T_0)$ operation to the Ward identity (13) with $n=3$, and making use of (24) (following again from gauge invariance) one obtains
\be
(T_1-T_0){\tilde\Gamma}^{(\Delta)a_1a_2a_3}_{\mu_1\mu_2\mu_3}
(k_1,k_2,k_3) = -2m(T_1-T_0){\tilde\Gamma}^{a_1a_2a_3}_{5;\mu_1\mu_2\mu_3}
(k_1,k_2,k_3)\ .
\ee
Since the LDD of the RHS is negative, its continuum limit is given by 
applying the $(T_1-T_0)$ operation to the integrand of the continuum box Feynman diagram, involving three vector vertices and a $\gamma_5$ insertion. After some work one finds that
\be
\lim_{a\to 0}{\tilde\Gamma}^{(\Delta)a_1a_2a_3}_{\mu_1\mu_2\mu_3}
(k_1,k_2,k_3) = i\frac{g^3}{4\pi^2}\epsilon_{\mu_1\mu_2\mu_3\sigma}f_{a_1a_2a_3}
q_\sigma 
\ee
where $q=k_1+k_2+k_3$. Finally, consider (30). Applying $T_0$ to the Ward 
identity (13) we obtain 
\be
\lim_{a\to 0}T_0{\tilde\Gamma}^{(\Delta)a_1a_2a_3a_4}_{\mu_1\mu_2\mu_3\mu_4}(k_1,k_2,k_3,k_4 ) = -2m \lim_{a\to 0}T_0 {\tilde\Gamma}^{a_1a_2a_3a_4}_{5;\mu_1\mu_2\mu_3\mu_4}(k_1,k_2,k_3,k_4 )\ .
\ee
Since the LDD of the vertex function on the RHS is again negative, the limit $a\to 0$ can be calculated by applying $T_0$ to the continuum pentagon graph involving a $\gamma_5$ insertion. A simple calculation shows that (35) is given by
\be
\lim_{a\to 0}{\tilde\Gamma}^{(\Delta)a_1a_2a_3a_4}_{\mu_1\mu_2\mu_3\mu_4}(k_1,k_2,k_3,k_4 ) =\frac{g^4}{4\pi^2}[\epsilon_{\mu_1\mu_2\mu_3\mu_4}Tr(T^{a_1}T^{a_2}T^{a_3}T^{a_4}) + perm.]
\ee
where $T^a$ are the $SU(3)$ generators in the fundamental representation. The (continuum) correlation functions in coordinate space, corresponding to 
(32), (34) and (36), are obtained 
by performing the Fourier transformation (11), where the limits of 
integration now extend to infinity, and with
${\hat\Gamma}^{(\Delta)a_1\cdot\cdot\cdot a_n}_{\mu_1,\cdot\cdot\cdot,\mu_n}$ 
defined in (12). The anomalous contribution is then given by (10) with ${\cal O} \to \Delta$, where the sum over the coordinates are replaced by 
integrals. Only the two and three 
gluon vertex functions actually contribute to the anomaly, as can be shown 
by making use of the Jacobi identity. One then easily verifies that the 
anomalous contribution to the (euclidean) axial vector Ward identity takes
the well known form
\be
{\cal A}(x) = -\frac{g^2}{32\pi^2}\epsilon_{\mu\nu\lambda\rho}F^a_{\mu\nu}(x)
F^a_{\lambda\rho}(x)
\ee
where 
\be
F^a_{\mu\nu}(x) = \partial_\mu A^a_\nu (x) - \partial_\nu A^a_\mu (x) 
-gf_{abc} A^b_\mu (x)A^c_\nu (x)
\ee
is the non-abelian field strength tensor. 

Finally, we point out that not only ${\tilde\Gamma}^{(\Delta)a_1\cdot
\cdot\cdot a_n}_{\mu_1\cdot\cdot\cdot\mu_n}$ possesses a finite limit, but 
also the other two vertex functions appearing in (13). The reason is, that the latter have LDD = 1 and 0 for $n=2$ and $n=3$, respectively, and negative LDD for $n\ge 4$. Gauge invariance, as embodied in the statements (23) and (24), 
allows us to replace the $n=2$ and $n=3$ vertex functions by the 
corresponding Taylor subtracted forms with negative LDD. These possess a 
finite continuum limit.
  
Concluding, we have shown, that any choice of gauge invariant lattice discretization 
of the action of QCD which is local in a more general sense and free of 
doublers will reproduce the correct axial anomaly. This anomaly, which arises from a naively irrelevant term in the Ward identity, is determined from the appropriate Taylor terms in the 
expansion of the triangle, box and pentagon continuum Feynman diagrams  
around vanishing momenta. Although we have explicitely referred to QCD, it is 
evident that the proof holds for any $SU(N)$ gauge theory.
\vskip7pt
\noindent
\centerline{\bf ACKNOWLEGMENT}
\vskip7pt
\noindent
We are very grateful to T. Reisz for several very enlightening discussions 
and constructive comments.

%%%%%%%%%%%%%%%%%%%%%%%%%%%%%%
%
%  Bibliography
%
%%%%%%%%%%%%%%%%%%%%%%%%%%%%%%

\end{document}